\documentclass[aps,showkeys,superscriptaddress,notitlepage]{revtex4-1}
\usepackage{graphics,graphicx} 
\usepackage{geometry}
\usepackage{epstopdf}
\usepackage{natbib}
\usepackage{enumitem}

\begin{document}

\title
{Tracking Urban Human Activity from Mobile Phone Calling Patterns}

\author{Daniel Monsivais}
\email[Corresponding author; ]{daniel.monsivais-velazquez@aalto.fi}
\author{Kunal Bhattacharya}
\author{Asim Ghosh}
\affiliation {Department of Computer Science, Aalto University School of Science, P.O. Box 15400, FI-00076 AALTO, Finland}
\author{Robin I.M. Dunbar}
\affiliation {Department of Experimental Psychology, University of Oxford, South Parks Rd, Oxford, OX1 3UD, United Kingdom}
\affiliation {Department of Computer Science, Aalto University School of Science, P.O. Box 15400, FI-00076 AALTO, Finland}
\author{Kimmo Kaski}
\affiliation {Department of Computer Science, Aalto University School of Science, P.O. Box 15400, FI-00076 AALTO, Finland}
\affiliation {Department of Experimental Psychology, University of Oxford, South Parks Rd, Oxford, OX1 3UD, United Kingdom}
\vspace*{0.2in}

\maketitle

\section*{Abstract}
Timings of human activities are marked by circadian clocks which in turn are entrained to different environmental signals. In an urban environment the presence of artificial lighting and various social cues tend to disrupt the natural entrainment with the sunlight. 
However, it is not completely understood to what extent this is the case.
Here we exploit the large-scale data analysis techniques to study the mobile phone calling activity of people in large cities to infer the dynamics of urban daily rhythms. From the calling patterns of about 1,000,000 users spread over different cities but lying inside the same time-zone, we show that the onset and termination of the calling activity synchronizes with the east-west progression of the sun. We also find that the onset and termination of the calling activity of users follows a yearly dynamics, varying across seasons, and that its timings are entrained to solar midnight. Furthermore, we show that the average mid-sleep time of people living in urban areas depends on the age and gender of each cohort as a result of biological and social factors.

\section*{Author Summary}
For humans living in urban areas, the modern daily life is very different from that of people who lived in ancient times, from which todays' societies evolved. Mainly due to the availability of artificial lighting, modern humans have been able to modify their natural daily cycles. In addition, social rules, like those related to work and schooling, tend to require specific schedules for the daily activities. However, it is not fully understood to what extent the seasonal changes in sunrise and sunset times and the length of daylight could influence the timings of these activities. In this study, we use a new approach to describe the dynamics of human resting periods in terms of mobile phone calling activity, showing that the onset and termination of the resting pattern of urban humans follow the east-west sun progression inside the same timezone. Also we find that the onset of the low calling activity period as well as its mid-time, are subjected to seasonal changes, following the same dynamics as solar midnight. Moreover, with  resting time measured as the low activity periods of people in cities, we discover significant behavioural differences between different age and gender cohorts. These findings suggest that the length and timings of the human daily rhythms, still have a sensitive dependence on the seasonal changes of the sunlight.


\section*{Introduction}
The daily activity of people varies across space and time from place to place, date to date, and hour to hour as a result of biological, societal, economic, and environmental factors, shaping the society where they live. Roughly speaking, each day humans do certain activities at specific times. There are many environmental factors (cues or `zeitgebers') involved in the entrainment of this clock, but as pointed out by Roenneberg et al. \cite{roenneberg2013light}, the most dominant is light and is associated with the light-darkness cycle determined by the daily rhythm of daylight. However, mainly in places not close to the equator, the timing and duration of daylight is subject to noticeable seasonal variation due to the yearly movement of the Earth around the Sun, and these changes have a direct influence on the kind and timing of different human activities. On the other hand, humans living in urban areas are also immersed in an environment full of cues that could influence the entrainment of the circadian clock. Artificial lighting, social practices and schedules (work and school hours, workdays vs weekends), particularly for those living in big urban areas, could have a noticeable influence on the entrainment process. Social conventions impose characteristic schedules on individuals, and, at the population level we can expect people in urban areas to have periods of high activity between morning and evening, and periods of low activity (resting) during the night. The length and timings of human activity periods, specifically in urban areas, has important consequences for human health \cite{dinges1995overview,pauley2004lighting,nutt2008sleep,hidalgo2009relationship}, economy and power consumption\cite{aries2008effect}, and public transportation efficiency \cite{sampaio2008efficiency}.


The human sleep wake cycle (SWC), and its dynamics in particular, has been studied in recent years to understand the processes and cues that govern it \cite{hofstra2008assess}. Generally speaking, most research on human SWC has focused on experiments with small groups under controlled conditions \cite{orzel2010consequences,davies2014effect}, or questionnaire studies\cite{roenneberg2004marker,roenneberg2007human,roenneberg2007epidemiology,levandovski2013chronotype,roenneberg2013light} (mainly using the Morning-Eveningness Questionnaire (MEQ)\cite{horne1975self} and the Munich Chronotype Questionnaire (MCTQ)\cite{roenneberg2003life}). The use of these tools for studying SWC has proved to be very fruitful and effective, though having some limits on the domain of applicability \cite{levandovski2013chronotype}. 
In contrast, the ever-increasing availability of information communication technologies (ICT) combined with researchers' ability to access large-scale ICT-generated datasets (`Big data') has made possible the study of human behaviour using a variety of reality (data) mining techniques. In particular, there are a number of examples where mobile phone datasets have been analyzed to study social networks \cite{kovanen2013temporal,eagle2009inferring,jiang2013calling,blondel2015survey}, sociobiology \cite{Bhattacharya160097,david2016communication}, mental health \cite{torous2015realizing}, mobility\cite{song2010limits,sevtsuk2010does,stopczynski2014measuring,jiang2016timegeo}, as well as social behaviour of cities \cite{sun2013understanding,louail2014mobile}. Over the past decade or so, the existence and accessibility of these large population-level datasets, has allowed scientists to study intrinsic human behavioural and socio-evolutionary patterns in unprecedented and complementary ways, compared to other research approaches.

Recently, datasets of mobile phone usage have also been used to study circadian rhythms, by analyzing individual's mobile phone usage from the  data captured by sensors \cite{stopczynski2014measuring,abdullah2014towards,aledavood2015daily,aledavood2016channel,monsivais2017seasonal,christensen2016direct,cuttone2017sensiblesleep}, or people's communication patterns from their call detail records (CDRs) \cite{aledavood2015daily,aledavood2016channel,monsivais2017seasonal}. For example, one study used the mobile phone screen on-off sensor data to examine the sleep wake cycle of nine individuals, finding that most of the individuals varied their sleep time patterns between weekdays and weekends, as well as showing seasonal changes in their mid-sleep time \cite{abdullah2014towards}. In another study using mobile phones calls and text messages of a small number of individuals, it was shown that individuals can be classified as having morning type or evening type activity levels \cite{aledavood2015daily}. In our previous related work \cite{monsivais2017seasonal}, we quantified the resting periods of people from their mobile phone calling activity, showing that there is a counterbalancing effect between the afternoon and night time resting periods, due to an interplay between ambient temperature and sunlight.
The use of CDRs as a tool for investigating the sleep/wake circadian rhythm, is in our view a promising new line of research as of that  complements the other research approaches especially the large scale survey-based studies, pioneered by Roenneberg et al \cite{roenneberg2004marker, roenneberg2007human,roenneberg2007epidemiology}. 

In this study, we apply reality mining techniques to users' call records in a mobile phone communication network to study the dynamics of the users' calling patterns by focusing on the periods of low activity, {\it i.e.} when almost no calls are made. Users of the mobile phone network typically have specific time periods during which their calling activity ceases, and we may assume that the SWC is bounded inside this period of inactivity. We observe that the daily calling activity time displays an interesting dynamics across the year through seasons and along different geographical zones. By studying these patterns we can gain insights into human activity patterns, and the SWC, in particular. Interestingly, the calling activity pattern changes with the day of the year and it is found to depend also on the geographical location  (latitude and longitude of the mobile phone user). From the circadian clocks involved in the daily rhythms of human societies, only those entrained to solar-based events depend also on the geographical location and on the day of the year.

In this work, we use mobile phone calling activity at the population level to study how the onset and termination of the urban human activity in different cities is synchronized with the East-West progression of the Sun. Also, we analyzed the annual progression of the onset and termination of the calling activity, finding that they show a strong seasonal variation. We note that this behavior is similar to the annual dynamics of solar midnight, inferring that solar midnight is an important cue entraining the human circadian clock. Finally, we determine the mid-time of the period of low calling activity, which is bounded between the termination of calling activity each day and its onset on the next day. We interpret  this mid-time to correspond to the mid-sleep time, and show that it is strongly dependent on the age and gender of the individuals in the population.

\begin{figure}[!ht]
\centering
\includegraphics[width=0.7\linewidth]{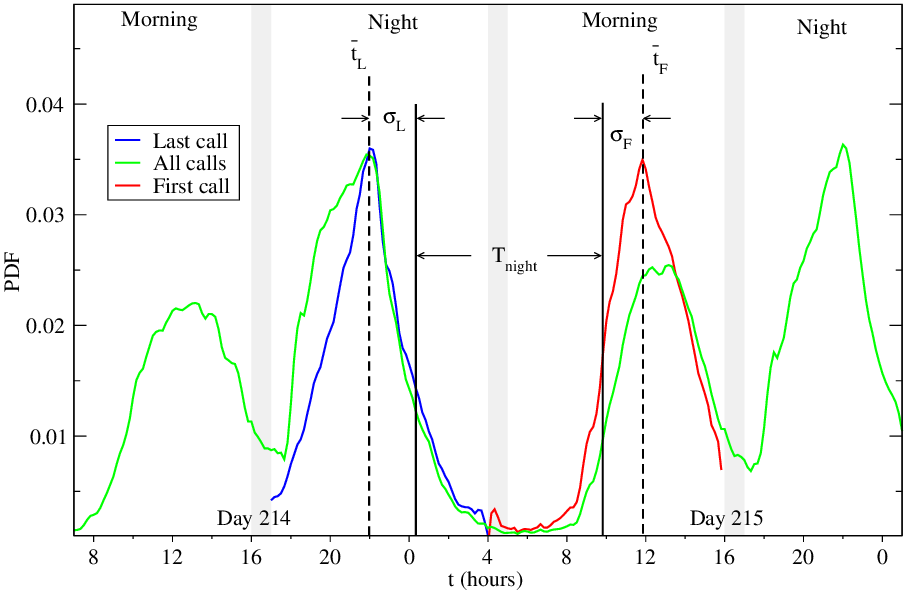}
\caption{{\bf Probability distribution for finding a call at time t, for a particular in 2007.} (green) Distribution when all the calls are included. (red) Distribution when only the last call at night is included (between 5pm and 4am next day).  (blue) Distribution when only the first call of the day is included (between 5am and 4pm). The distribution of the last and first calls are sharper and have well-defined maxima.}
\label{Fig1}
\end{figure}

\section*{Results}
Using an anonymized dataset containing details of mobile phone communication of subscribers of a particular operator in a European country described in detail in the Methods section, we investigate the calling activity of the urban population living in cities as a function of time of the day for all the dates during the year. This we do by calculating for each city the probability distribution $P_{all}(t,d)$ for finding an outgoing call at time $t$ of a day $d = (1,...,365)$ of the year. For all the studied cities, a region of almost null activity can be found around 4:00 am. Using this natural bound to split the calling activity from one day to another, we define a `day' starting from 4:00am of a calendar day and running to 3:59am of the next calendar day.

In Fig.~\ref{Fig1} we show $P_{all}(t,d)$ (green line) during days $d$$=$$214-215$ (marking early August) for a city with over a 500,000 inhabitants. The distribution $P_{all}(t,d)$ has two high calling activity periods with the first one corresponding to the morning calls, peaking around noon, and the second related to the evening calls, peaking around 8:00 pm. This bimodal pattern is present every day across the year and all the cities included in this study. The high calling activity periods are delimited by two periods of low activity, one centered around 4:00 pm related to the time after lunch, and the second one in the middle of the night, around 4:00 am within the sleeping period. The pattern present in $P_{all}(t,d)$ is similar to that reported in other studies using different CDRs \cite{louail2014mobile,mollgaard2016general}, mainly at the times when the calling activity starts and ceases. In ref. \cite{louail2014mobile}, where the calling activity of some Spanish cities was studied, the histogram of the number of active users at each time has a similar bimodal shape, with similar times for their onset and termination, as well as the depth in the middle located around the same time period (i.e. between 3:00 pm and 4:00 pm).

To study the specific times when the calling activity rises and falls, we analyze the `morning' and `night' periods  separately, defining the former  between 5:00 am and 3:59 pm, and latter between 5:00 pm and 3:59 am on the following calendar day, in such a way that each period is 11 hours long. During each `morning', we select only the first call made by each user inside that period and construct the associated probability distribution for the time of the first call $P_F(t,d)$, directly related to the rise of calling activity.  Similarly, during the `night' we define the corresponding probability distribution for the time of the last call $P_L(t,d)$ by taking into account only the last call made by each user within that period. In Fig.~\ref{Fig1}, it can be seen that the three defined probability distributions $P_{all}(t,d)$ (green), $P_L(t,d)$ (red), and $P_F(t,d)$ (blue) for consecutive days during winter, for a particular city with a population over a 500,000. The shape of the distributions $P_{all}(t,d)$, $P_L(t,d)$, and $P_F(t,d)$ depicted in Fig.~\ref{Fig1} for a specific day appear to be preserved for all the days and cities we have studied.

\subsection*{Urban activity synchronization with East-West Sun progression}
\begin{figure}[h!]
\centering
\includegraphics[width=0.7\linewidth]{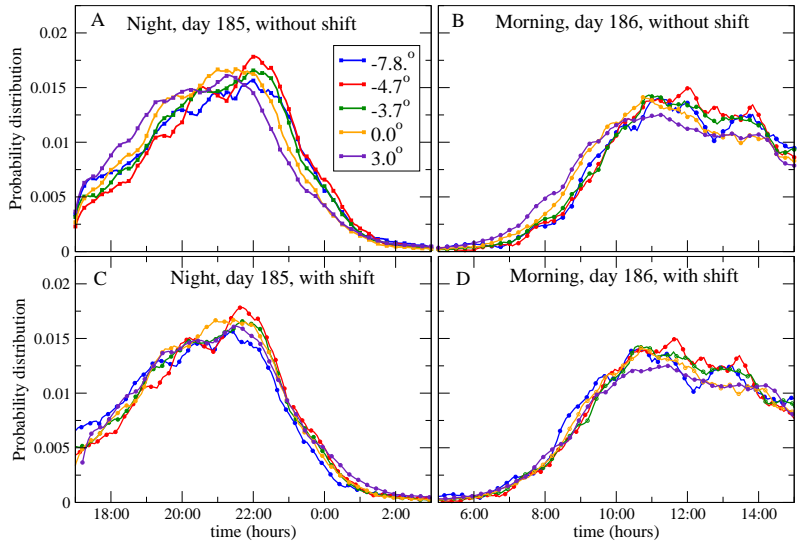}
\caption{{\bf Temporal shift of the onset and termination times of the calling activity along geographical longitude. }Probability distributions of the time of the last call $P_L(t,d)$ and that of the first call $P_F(t,d)$ for 5 different cities lying at the same latitude but at different relative longitudes from a reference point located at the second city from east to west within the band for two consecutive days during the year. The relative longitudes of the cities are -7.8$^{\circ}$, -4.7$^{\circ}$, -3.7$^{\circ}$,  0$^{\circ}$, and 3$^{\circ}$. (Upper panel) Probability distributions for  (A) the time of the last call, and (B) the time of first call. (Lower panel) Probability distributions for (C) the time of the last call, and (D) the time of first call, shifted by a time corresponding to the difference between their local sun transit times (31.2, 18.8, 14.8, and -12 minutes for the cities located at -7.8$^{\circ}$, -4.7$^{\circ}$, -3.7$^{\circ}$, and 3$^{\circ}$ from the reference city, respectively).  The collapse of the distributions onto the reference city's distribution is evident when the longitudinal time shift is added. This collapse implies that the 5 cities begin (or cease) their calling activity in a way that is synchronized with a temporal phase corresponding to the difference between their sun transit times.}
\label{Fig2}
\end{figure}

The mean time of the first call $t_F$ and of the last call $t_L$ of people in a city can be influenced by environmental, social, and economic factors, and their possible daily value could be distributed completely at random. However, we find that during the year and at different latitudes, despite the different factors influencing the shape of the distribution $P_{all}$, the onset and termination of calling activity follows a consistent pattern, and this characteristic behaviour allows us to compare the calling activity pattern of cities lying at different latitudes. If the onset or termination of the urban calling activity is socially driven, with fixed times for specific activities (like office working hours from 9:00am to 6:00pm), one could expect that cities lying in the same time zone and at the same latitudes have similar calling activity timings (onset and termination). However, we find that the onset and termination of calling activity synchronizes with the East-West sun progression, in such a way that cities lying in western locations start (and terminate) their calling activity after cities at eastern locations, with a delay difference corresponding to the time difference between their local meridians. In Figs. \ref{Fig2}A-B we show $t_L$ and $t_F$ for 5 different cities lying inside a latitudinal band centered at $42 ^ \circ$N$\,\pm 40'$. The region including the 5 cities spans a longitudinal angle of $10.8^\circ$, and by taking one of the cities as a reference, other cities are located at $-7.8^\circ$, $-4.7^\circ$, $-3.7^\circ$, and  $3.0^\circ$  from the reference city marked here with $0.0^\circ$. Then we compare the actual distributions $P_L$ and $P_F$ of the time of the last call and of the first call, respectively, for the 5 cities in the same latitudinal band, and find that $P_L$ and $P_F$ for western cities seem shifted to later times. 
However, when the distributions are shifted by an amount of time corresponding exactly with the time difference between the local meridian of the corresponding city and the reference city, the distributions visibly collapse onto each other, as can be seen in Figs. \ref{Fig2}C-D. In this case, the time shifts are +31.2, +18.8, +14.8, and -12 minutes for the cities located at -7.7$^{\circ}$, -4.7$^{\circ}$, -3.7$^{\circ}$, and +3$^{\circ}$ from the reference city at 0$^{\circ}$, respectively.

\begin{figure}[!ht]
\centering
\includegraphics[width=.7\linewidth]{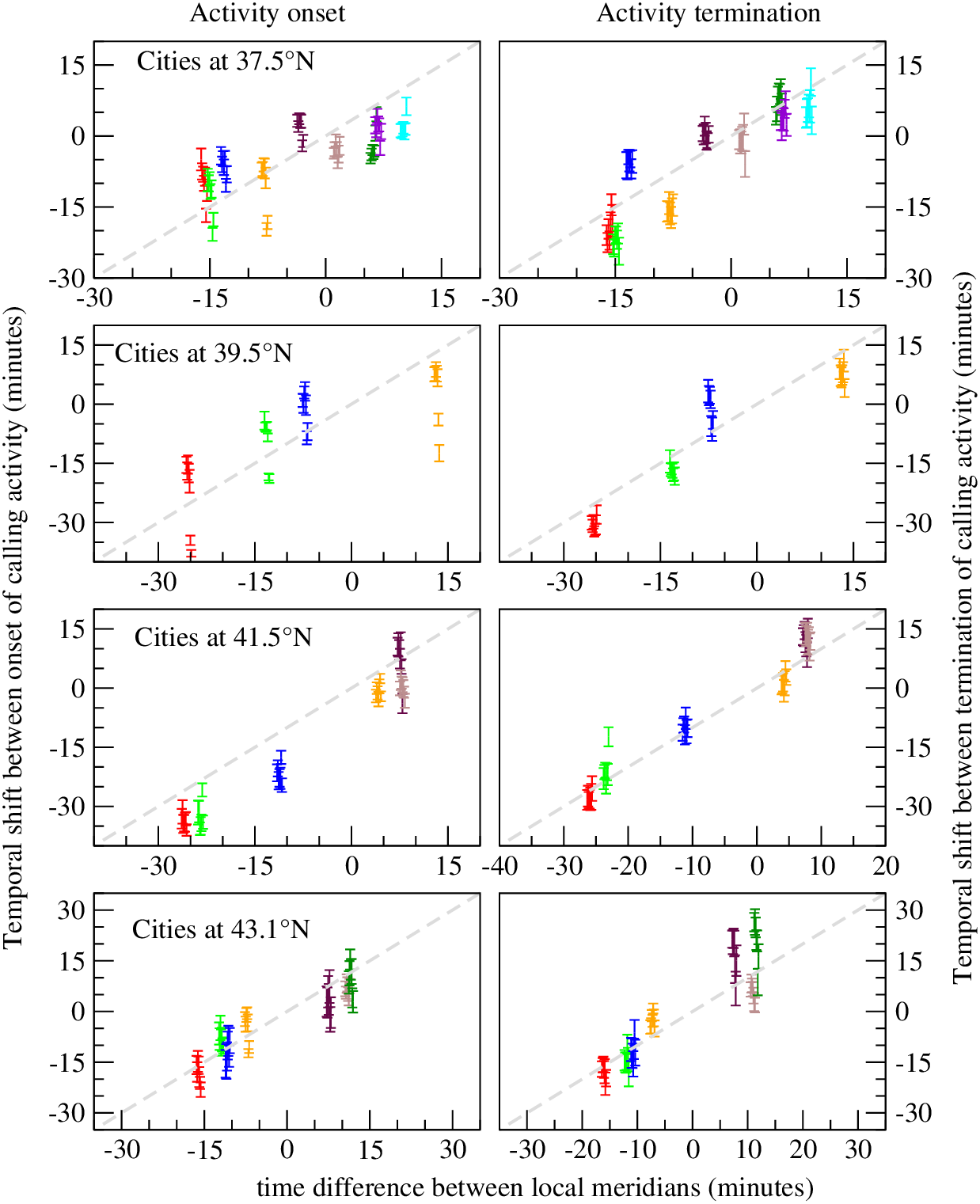}
\caption{
Temporal progression of the onset and termination of the calling activity for cities lying at different geographical longitude}.The time shift $n^*\Delta$ that minimizes the divergence between the probability distribution of the first call $P_F$ in a reference city and the corresponding distributions of the other different cities lying at the same latitude. 4 different bands are analyzed, centred at $37.5^\circ$N, $39.5^\circ$N,  $41.5^\circ$, and $-43.1^\circ$. For each city inside each band, the time shifts  $n^*\Delta$ for the 7 days of the week are shown, as the set of 7 points with the same color located at the corresponding time difference between the local meridians of each city and that of the reference. The dashed line represents the time shift between the sun transit time at the reference city and a hypothetical point located at each corresponding longitude. The error bars represent the standard deviation from the average value for each day of the week. From the plot it can be seen that, for cities lying further away from the reference city, a bigger time shift is required to collapse the distributions.
\label{Fig3}
\end{figure}

The distribution collapse shown in  Figs. \ref{Fig2} is obtained by introducing a time shift corresponding to the sun transit differences between cities. In order to quantify the exact delay between the distributions, we calculate the required time shift that should be introduced between the calling distributions to minimize the Kullback-Leibler divergence $D_{KL}$ between them (see the Methods section). This measure is indicative of the similarity between the distributions, and is minimized when they are identical. We  extend this analysis to include data from 30 cities, each one lying in one of the four latitudinal bands centered at $37 ^ \circ$N (10 cities), $39.5 ^ \circ$N (5 cities), $41.5 ^ \circ$N  (7 cities), and  $42.5 ^ \circ$N (8 cities). For each band, we choose one city lying near the mid point of the band as the reference, and calculate for all the cities in the band the average time shift between them and the reference city. This is done for each day of the week, averaging over 52 weeks of the year 2007. The results are shown in Fig. \ref{Fig3}, and it can be seen that the time shift that minimizes the divergence between the distributions corresponds to the delay between their local sun transit times. This synchronization appears stronger for the termination of the calling activity (represented  by the distributions $P_L$). As this pattern is consistently present in all of the four analyzed latitudinal bands, we conclude that it is a general behaviour of the population living in the cities. This result is consistent with those reported by Roenneberg {\sl et al.}\cite{roenneberg2007human}, obtained from MCTQ studies of people in Germany, distributed over a region that is $9^\circ$ wide longitudinally. In their work, they take into account the population of the city by defining three population size categories, i.e. less than 300,000  inhabitants, between 300,000 and 500,000 inhabitants, and more than 500,000 inhabitants, while we classify each city of more than 100,000 inhabitants according to its latitudinal coordinate. Grouping the cities into latitudinal bands, we found a consistent entrainment to the East-West progression of the Sun, regardless of the population size of each city.

This result implies that the termination (last call of the day) and onset (first call of the next day) of calling activities in cities at similar latitudes follow an external cue driven by solar events, and the time difference in these solar events between two different cities is reflected in the timings of their calling activity. 

\subsection*{Entrainment of urban calling activity with sun-based cues}
We have shown that the cities located at the same latitude but at different longitudes have periods of low calling activity with different onset and termination times (Figs. \ref{Fig2} and \ref{Fig3}). This shift coincides with the difference between their local sun transit times, i.e. when the sun crosses the meridian of the city. This observation raises the question as to what external daily event induces such synchronization. As the delays correspond to the time period between the local sun transit times of the cities, it seems plausible to think that the sun functions as a cue for this entrainment.

At the latitudes where the studied cities are located, the time difference between the sunset in the summer and in the winter is around 3 hours, if daylight saving is not taken into account, and the same holds for the time difference between sunrises. In contrast, the time difference between the mean time of the last calls between summer and winter is at most one hour \cite{monsivais2017seasonal}. However, there is a clear synchronization between the sun transit time and the timings of calling activity. This means that there should be an external clock functioning as a cue. On the other hand, from a biological perspective, the time when the secretion of melatonin reaches its maximum\cite{duffy1999relationship} lies close to midpoint between sunset and sunrise ({\it i.e.} solar midnight), once the night is as dark as possible. It has been proposed that the mid-sleep time coincides with the time corresponding to maximum melatonin secretion\cite{dijk1997variation,dijk2012amplitude}, and if the solar midnight shifts through the year, the time for the maximum melatonin secretion should follow a similar pattern, as well as the entrained mid-sleep time. 

In their study, Allenbradt {\it et al.}\cite{allebrandt2014chronotype}, using the MCTQ approach, have reported that mid-sleep time (on free-days) changes from one season to another. In some of the studied populations, they found that  there is a small but significant difference in the average mid-sleep time between the days when Daylight Saving Time is applied and other days. This lends support to our assumption that if the mid-sleep time shifts in response to seasons, the timings of the calling activity should be influenced by its variation. In such a case, when the human mid-sleep time occurs at later hours, the timings of the calling activity for the following days should also occur at later hours. In other seasons, when the mid-sleep time occurs earlier, the activity timings should also be shifted towards earlier hours. If this is the case, then solar midnight should be functioning as the cue to which the calling activity timings are entrained. The activity pattern is a consequence of the interplay between seasonal and geographical factors, as well as social and societal activities like work and/or school, transportation, eating and leisure activities. However, the latter require specific timings during the day, not necessarily controlled by the sleep/wake cycle. We have shown elsewhere  \cite{monsivais2017seasonal} that the total period of low calling activity (that is, the period between the termination and the onset of the calling activity) is strongly correlated with the duration of daylight, showing seasonal changes similar to the mid-sleep time.
\begin{figure}[!ht]
\centering
\includegraphics[width=.7\linewidth]{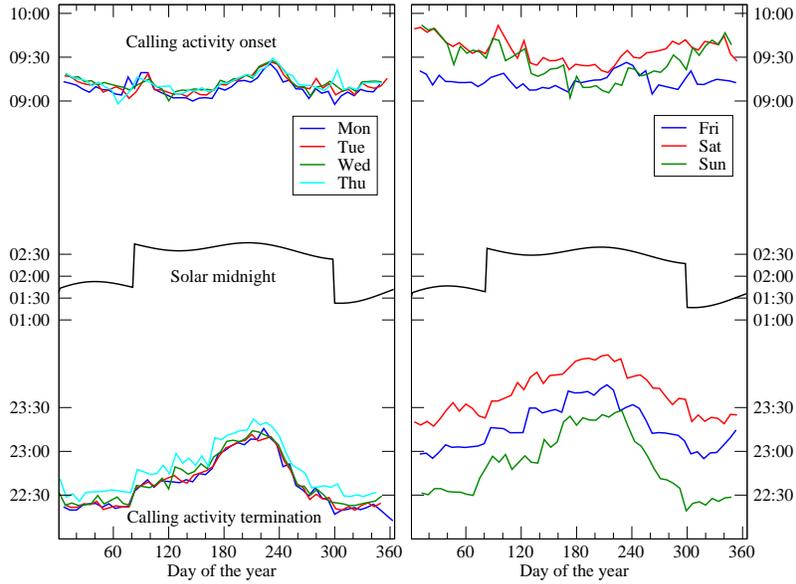}
\caption{{\bf The yearly evolution of the time of the first call and that of the last call compared against the yearly shift of the solar midnight}. (Top sets) ${\overline{t}_F}$ -- average of the  mean time of the first call of 3 sets of cities located at latitudinal bands centred at $\phi=37^\circ 30'$N (blue), $40^\circ 20'$N (green), and $43^\circ 0'$N (red). (bottom sets) ${\overline{t}_L}$ -- average of the mean time of the last call for the same sets of cities. In the middle of the panels, the solar midnight time in one of the cities within the band. The shape of $\overline{t}_L^\phi$ resembles to some extent the graph of the solar midnight, coinciding with the two minima (for days 130 and 302) and one of the maxima (for day 210). For the case of $\overline{t}_{F}^\phi$, the graph shows some correspondence with the sunrise although to a lesser extent.  The discontinuities introduced by the daylight saving shows in the graphs, suggesting that the period of low calling activity is not solely influenced by the socially-driven time, but is synchronized with an external (astronomical) event. The number of cities inside the bands $\phi=37^\circ 30'$N (blue), $40^\circ 20'$N (green), and $43^\circ 0'$N (red), are  7, 6, and 8, respectively.}  
\label{Fig4}
\end{figure}

In order to find any possible synchronization between the onset (and termination) of calling activity and solar midnight, we calculate the average of the mean times of the last call ${\overline{t}_L}$ and that of the first call ${\overline{t}_F}$, for three sets of cities located at the latitudinal bands $\phi=37^\circ 30'$N (seven cities), $40^\circ 20'$N (six cities), and $43^\circ 0'$N (eight cities). We compare ${\overline{t}_L}$, and ${\overline{t}_F}$ with the yearly evolution of the solar midnight in a reference city within a given latitudinal band (see Fig. \ref{Fig4}). A detailed description of how  ${\overline{t}_L}$ and  ${\overline{t}_F}$ are calculated can be found in the Methods section. It can be seen that only $\overline{t}_L$ resembles to some extent the dynamics of the solar midnight, with their two minima and at least one of their maxima occurring around the same days of those of solar midnight, although the relative amplitudes are not in correspondence. In addition, the discontinuities introduced by the daylight saving is visible in all the graphs, suggesting that the timings of the calling activity are not solely influenced by the socially-driven time, but instead are synchronized with an external (astronomical) clock.  

\subsection*{Age and gender dependence of the mid-sleep times }

The period of low calling activity is bounded by the mean times of the last call during the night and of the first call in the morning. The duration of this period changes across seasons \cite{monsivais2017seasonal} and is strongly influenced by the length of the day (or conversely by the length of the night). 
 The mid-time of this low calling activity period should correspond to the average time of human low activity, i.e. when the majority of the urban population is sleeping. In chronobiology studies, the mid-sleep time, corresponding to the time when human sleep is in the middle of its cycle, has been found to vary with the age and gender of the individuals\cite{roenneberg2004marker,foster2008human}. Despite the fact that each individual has a distinctive sleep-wake cycle, with a chronotype ranging from advanced sleep period (morningness) to delayed sleep period (eveningness) \cite{lockley2012sleep}, at the population level a characteristic mid-sleep time can be consistently calculated, taken simply as the average of individual mid-sleep times. 

\begin{figure}[!ht]
\centering
\includegraphics[width=.7\linewidth]{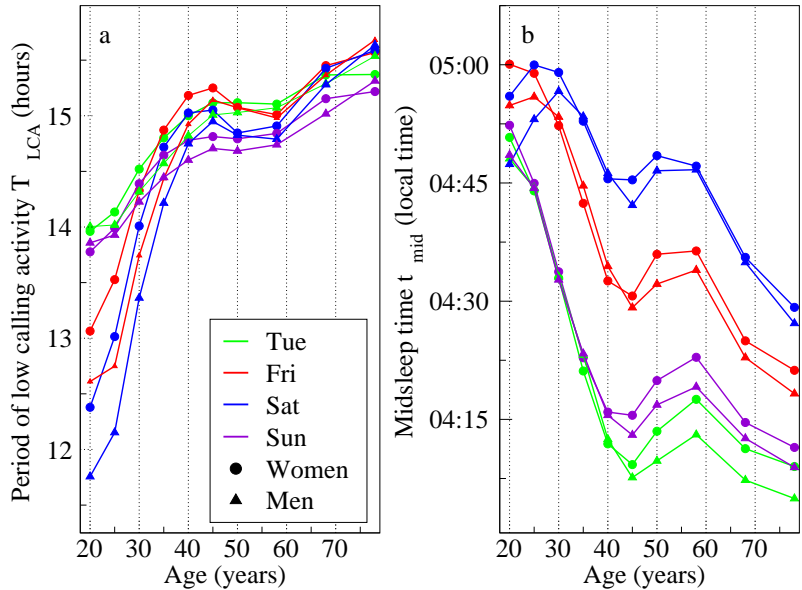}
\caption{{\bf Period of low  calling activity and mid-sleep times for different age and gender cohorts} (a) Period of low  calling activity $T_{LCA}$. The $T_{LCA}$ is calculated as the elapsed time between the mean time of the last call and that of the first call, as a function of the age and gender of different cohorts, for the most populated city in the dataset in 2007. (b) mid-sleep time  $t_{mid}$, calculated as the time in the middle of the interval between the mean time of the last call and that of the first call, as a function of the age and gender of different cohorts of the same city. For each age cohort, $T_{LCA}$ and  $t_{mid}$ are calculated for females (circles) and males (triangles) separately. Both quantities are different for different days of the week, and the corresponding plots are shown for (green) Tuesdays, (red) Fridays, (blue) Saturdays, and (violet) Sundays. As Mondays to Thursdays have similar values, therefore only the data for Tuesdays is shown.}  
\label{Fig5}
\end{figure}

From the mean times of the last call of the day, $t_L$ and of the first call $t_F$ of the next day, we define the period of low calling activity $T_{LCA}$ as the elapsed time between $t_L$ and $t_F$, as a measure of the time when cities cease their activity. In Fig. \ref{Fig5}a, the width of the low activity period $T_{LCA}$ of the most populated city in the dataset is shown, for 4 different days of the week (Tuesdays, Fridays, Saturdays and Sundays), as a function of the subscribers' age and gender. There is a noticeable change of about 3 hours, moving from the age cohort of 20 to that of 40 year olds. After that rather abrupt increase, especially for Fridays and Saturdays, $T_{LCA}$ slightly decreases, reaching a local minimum value for the age cohort of 50 year olds, and then it increases again to reach the highest value at the age of 78 years. For the analyzed weekday (Tuesday) as well as for Sunday, $T_{LCA}$ increases almost monotonically with the cohort age, showing a small plateau for age cohorts between 45 and 58. 

We have also tracked the midpoint of the inactivity period, defined as the mid-time between $t_L$ and $t_F$. Due to its similarity with the average time in the middle of the sleeping period \cite{foster2008human}, we interpret this minimum calling activity time as the mid-sleep time $t_{mid}$, calculated simply as $t_{mid}=(t_L+t_F-24)/2$. Both quantities are found to depend on the age and gender of each cohort, as can be seen in Fig. \ref{Fig5}b. We find that, for certain age groups (from 18 to 32 years old, and from 43 to 80 years old)  $t_{mid}$ occurs at a later time for women as compared to men, while in the age group of 33 to 42 years old, $t_{mid}$ for the men occur later. This finding differs somewhat from the reported mid-sleep times (on free days) in the chronotype questionnaire study  based on the MCTQ  \cite{roenneberg2004marker,roenneberg2007epidemiology}, where males show a later mid-sleep time for age cohorts younger than 38 years old. Also, there is a strong dependence on age, with younger age cohorts (20--30 year old) having later $t_{mid}$, i.e. around $30$ minutes after that of the oldest age cohort (70-80 years old). This observation is in accordance with the observed chronotypes \cite{foster2008human}, which are attributed to biological factors or internal clock being regulated by neuronal and hormonal mechanisms. We also found an unexpected rise of $t_{mid}$ for the age cohort of $45$--$65$ year old individuals, which we  suspect is entirely of social origin. Hence it seems that both biological and social factors play a role in changing $t_{mid}$, i.e. shifting the period of low activity to later hours. 

In addition, we find that $t_{mid}$ varies across days of the week. On Fridays and Saturdays $t_{mid}$ occurs at a later hours compared with the other days. Similarly, the age cohort with the latest mid-sleep time $t_{mid}$ is different for different days of the week. On Saturdays,  individuals in the age group 30 to 45 years old have the latest $t_{mid}$, while for the other days of the week it is the 20--25 years old cohort which shows the latest mid-sleep time. The results of $T_{LCA}$ and $t_{mid}$ for the most populated city are also and consistently found in the next 5 most populated cities, as shown in the Supplementary Information (Figs. \ref{SI-Fig1} and \ref{SI-Fig2} respectively).

\section*{Conclusion}
In this study, we have found that the onset and termination of the period of low calling activity for people in cities at about the same latitude but at different longitudes are shifted according to their relative longitudinal separation. Cities westward from the easternmost analyzed city stop their activity later in line with the time delay of the sun transit time. This result suggests that a solar event acts as a cue for the circadian rhythm of the period of low calling activity with the SWC bounded inside. 
This result is consistent with those reported by Roenneberg {\it et al.}\cite{roenneberg2007human}, although strictly speaking the two studies cannot be compared directly as the focus of our study is on variation by latitude and theirs was on variation by population size of cities.

In addition, we found that the seasonal variation of the termination of  calling activity resembles the annual variation in solar midnight (or solar noon). However, when the annual behaviour of activity termination  is compared with other characteristic solar events like the sunrise and sunset, it appears to have a different functional form with different number of maxima and minima  with different dates. Although, it seems likely that solar midnight (or solar noon) acts as a cue in the synchronization of the termination of the calling activity, further research is needed to confirm this.  At the individual level, knowledge of the mid-sleep time and sleep duration allows the determination an individual's chronotype \cite{roenneberg2003life}. However, at the population level, we could determine from the calling distributions the characteristic variation in the sleep duration and mid-sleep time as a function of the group age.
 The observed overall trends are in line with the earlier findings \cite{foster2008human} and reveal an increase in the sleep duration and decline in the mid-sleep time with age. Several other intricacies are also evidenced at closer inspection. Firstly, the aspect of `social jetlag' \cite{wittmann2006social}, defined as the difference between the mid-sleep times on free days and that of work days, becomes apparent across all age groups. Interestingly, although social jet-lag is expected to give rise to extended sleep duration on free days as a compensatory effect, for young adults ($20$--$25$) we find that the sleeping periods are comparatively less on free days (Friday and Saturday nights and the following mornings). Therefore, sleep deprivation is likely to be at a maximum for this age range. Second, previous observations suggest a monotonic decrease in the mid-sleep time from around $20$ years of age, which can be attributed to endocrine factors \cite{foster2008human}. In contrast, we observe a reversal in trend of the mid-sleep time such that at the age of $45$ years it starts rising till $55$ years of age, after which it decrease again. 
\section*{Materials and Methods} 
In this study, we have analyzed a very large dataset of anonymized call detail records (CDRs) from a mobile phone service provider offering services in a a country located in the Southern Europe subregion of the United Nations geoscheme \cite{ungeo}.  Due to a Non Disclosure Agreement associated to the dataset we are bound to 
keep the identity of the country unknown, and thus we have partially masked the latitude and Longitude coordinates of the cities to screen their actual location, such that each 
city is associated with a latitudinal band, and the latitude at the center of the band is assigned as the latitude of the city. In the analyses, depending on the measure we were focusing on, we chose the width and center of the latitudinal bands and in all cases specifying the corresponding values. The latitude coordinate associated with each band is described by $\phi \pm d\phi$, with $\phi$ the latitude in degrees at the center of the band, and $d\phi$ the half-width of the band in degrees. On the other hand, as the latitudinal region is given, the Longitude coordinate is also screened 
by providing instead its angular separation from, an arbitrary point located in the same latitudinal band. Thus, for a given city, its longitude coordinate $\theta$ denotes the number of degrees from a reference point located in the same latitudinal band. The anonymization of the subscribers' identities was performed by the service provider prior the data been given to us. The dataset contains CDRs of around 10,000,000 subscribers during 2007, with more than $3$ billion calls between 50,000,000 unique identifiers. Each record contains the date, time, duration, and anonymized caller and callee identifiers. The dataset also includes demographic information of the majority of the subscribers, and, for those cases, the age, gender, postal code, and location of the most accessed cell tower (MAC-tower) are known. Thus, there are three possible locations associated to each user, namely the associated city center, the location of the MAC-tower and the center of the postal code region, and we use them to determine whether the subscriber ``lives in a city'' -- defined by cases where their three associated locations are sufficiently close to each others.
  Taking as a reference point the geographical location of the associated city center, a subscriber lives there if the following three conditions are satisfied:
\begin{enumerate}[label=(\roman*)] 
\item the distance between the location of the MAC-tower and the associated city's center is less than 15 km
\item the distance between the center of the associated postal code region and the associated city's center is  less than 15 km
\item the distance between the location of MAC-tower and the center of associated postal code region is less than 30 km.
\end{enumerate}

In this study, we chose 36 of the cities with more than 100,000 
inhabitants in 2007, in such a way that our final analysis takes into account the calling patterns of around 1,000,000 subscribers in total. 
 Locations of the subscribers are associated with the locations of the cities they reside. Each city is associated with the following two geographical coordinates: the latitudinal coordinate is fixed as the midpoint of a latitudinal band including the city, and the longitudinal coordinate, defined as the angular distance between the city and a reference point located in the studied region.

\subsection*{Quantifying delays between calling activity timings}
Calling behavior varies seasonally, particularly the mean value and the width of the distributions of the first and last call vary across the year, being pushed towards the afternoon during winter and towards midnight during the summer. In spite of this seasonal variation, for a given day the calling distributions of different cities have similar shapes, and we exploit this similarity to calculate the delays between them to identify the temporal shifts of the distributions. 
The Kullback-Leibler divergence\cite{kullback1951information} is a measure of similarity between two distributions, commonly used in statistical analysis, for example when comparing one distribution obtained from data and another generated by a model. It reaches zero, its minimum possible value, when the distributions are identical, and it increases in value as the distributions become more and more dissimilar. In the case of the calling activity of different cities, the distributions are not identical but have a very similar shape. Applying Kullback-Leibler divergence to a pair of these distributions, it would reach a minimum value when these distributions overlap most, falling on top of each other and collapse to one. Thus, if we measure the amount of time one distribution should be shifted in order to minimize its divergence from the second distribution. The time shift would correspond the actual time delay between them.

In order to quantify the actual time shift between the distributions $P_L$ of last calls for cities lying along different Longitudes, we proceed as follows. First, for all the cities within the band, we calculate all the distributions $P_L(t,d)$ between January 2nd and December 31st. For each day $d$, we fix  $P_L(t,d)_{0^\circ}$ of the city labeled `$0^\circ$' as the reference distribution, and for every other city $c$ in the band, we compared the reference $P_L(t,d)_{0^\circ}$ with time-shifted versions $P_L\left(t+n\Delta,d\right)_{c}$ of the distribution $P_L(t,d)_{c}$,  with $-5\le n \le 8$ and $\Delta=5$ min, to find the time shift $n^*\Delta$  that minimizes the divergence $D_{KL}$ between them. Here, $D_{KL}$ is the Kullback-Leibler divergence measure, defined as $D_{KL}(P,Q)=\sum_i P_i \log\left(P_i/Q_i\right)$, with $P$, $Q$ being the two discrete distributions. Once we find 
 for each city the set $\lbrace n^*\Delta \rbrace$ with all the time-shifts across the year, we calculate its average time-shift $\left<n^*\Delta\right>$,  and plot it for all the cities in the band in the right column of Fig.~\ref{Fig3}. As the time for the mean time of the last call is different for different days of the week \cite{monsivais2017seasonal}, the average is calculated separately for each day of the week.  We apply the same procedure for the time of the first call distributions $P_F$, and the results are shown in the left column of  Fig.~\ref{Fig3}. 

\subsection*{Averaging the mean times of the calling activity inside a latitudinal band}

In order to find if there is any relation between $t_L$ and $t_F$ and the solar midnight, we have chosen $7$, $6$ and $8$ cities, lying in the latitudinal bands centered at $\phi=37^\circ 30'$ N, $40^\circ 20'$ N, and $43^\circ 0'$ N, respectively. For each city, we shift its corresponding distributions in accordance with its longitudinal difference to collapse all into one. Then we calculate the average mean time of the last call, $\overline{t}_L(d)=\langle\overline{t}'_L(d,c)\rangle$, where, $\overline{t}'_L(d,c)$ denotes the mean time of the last call for the shifted distribution for a city $c$ belonging to the analyzed band during the day $d$, and $\langle\cdot\rangle$ denotes the average over all cities lying within the band. Similarly, we calculate the average mean time of the first call $\overline{t}_F(d)$ for the given latitudinal band. The quantities $\overline{t}_L(d)$ and $\overline{t}_F(d)$ are compared with the time at which the solar midnight occurs in the reference city of each band. It should be noted that in the original graphs there are days of national holidays and local festivities that introduce drastic pattern changes, which we filter out to construct the final graphs.

\section*{Acknowledgments}
AG and KK acknowledge project COSDYN, Academy of Finland (Project No. 276439) for financial support. DM, KB, and KK acknowledge EU HORIZON 2020 FET Open RIA project (IBSEN) No. 662725. DM also acknowledges CONACYT (Mexico), grant No. 383907. RD acknowledges European Research Council for the Advanced Investigator Grant No. 295663.
\section*{Competing interests}
The authors declare that they have no competing interests

\clearpage

\section*{Acknowledgements}
AG and KK acknowledge project COSDYN, Academy of Finland (Project No. 276439) for financial support. DM, KB, and KK acknowledge EU HORIZON 2020 FET Open RIA project (IBSEN) No. 662725. DM also acknowledges CONACYT (Mexico), grant No. 383907. RD acknowledges European Research Council for the Advanced Investigator Grant No. 295663. Authors thank AL Barab\'asi for the dataset used in this research.


\section*{Additional information}

\subsection*{Competing financial interests} 
The authors declare no competing financial interests.
\section*{Author contributions statement}
K.B., A.G. and D.M. carried out the analysis of the data. All the authors were involved in designing the project and the preparation of the manuscript.
\clearpage
\appendix

\clearpage
\section*{Supporting Information}

\begin{figure}[!ht]
\centering
\includegraphics[width=.6\linewidth]{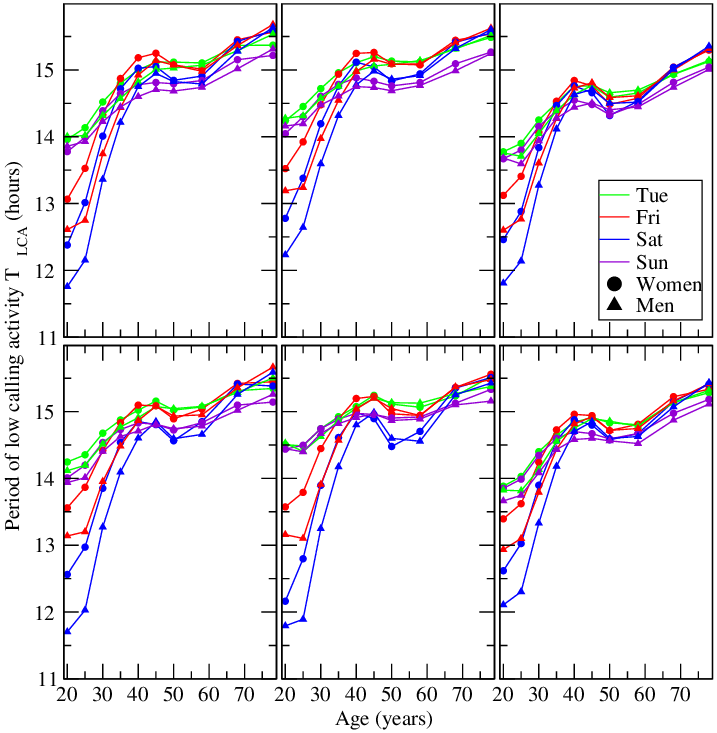}
\caption{ {\bf Period of low calling activity $T_{LCA}$ for different age and gender cohorts.} The $T_{LCA}$ is calculated as the elapsed time between the mean time of the last call and of the first call, as a function of the age and gender of different cohorts, for the six most populated city in the dataset in 2007. For each age cohort, $T_{LCA}$ is calculated for females (circles) and males (triangles) separately. $T_{LCA}$ is different for different days of the week, and the corresponding plots are shown for (green) Tuesdays, (red) Fridays, (blue) Saturdays, and (violet) Sundays. Mondays to Thursdays have similar values, therefore only the data for Tuesdays is shown.}  
\label{SI-Fig1} 
\end{figure}

\begin{figure}[!ht]
\centering
\includegraphics[width=.6\linewidth]{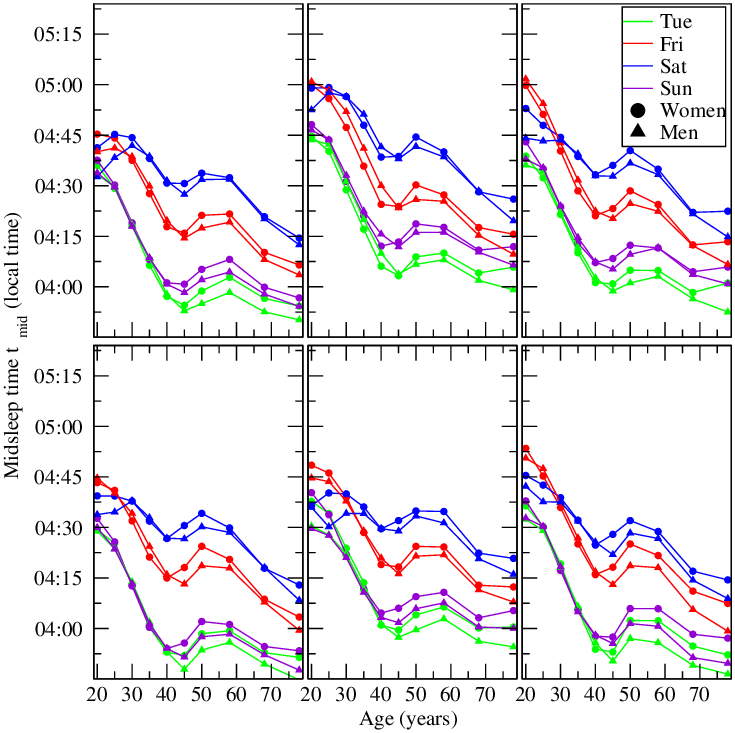}
\caption{{\bf mid-sleep time  $t_{mid}$ for different age and gender cohorts.} $t_{mid}$ is calculated as the time at middle of the interval between the mean time of the last call and of the first call, as a function of the age and gender of different cohorts, for six of the seven most populated cities in the dataset in 2007. For each age cohort, $t_{mid}$ is calculated for females (circles) and males (triangles) separately. $t_{mid}$ is different for different days of the week, and the corresponding plots are shown for (green) Tuesdays, (red) Fridays, (blue) Saturdays, and (violet) Sundays. Mondays to Thursdays have similar values, therefore only the data for Tuesdays is shown. During some Fridays nights, the calling activity extended until very late in the night, and the distribution of the morning calling activity on the next day presents a small peak around 4:00 a.m. If present, we include this peak in the analysis when calculating the time of the first call, due to its small amplitude and width compared with the main part of the distribution for the time of the first call. This is also true for the results shown in Fig. \ref{Fig5} in the main text.}  
\label{SI-Fig2}
\end{figure}

\paragraph*{S1. }
\label{S1_Fig}
{\bf Age and gender dependence of the period of low calling activity and  of the mid-sleep times.}
The behaviour of $T_{LCA}$ and $t_{mid}$ as a function of age and gender is calculated for the most populated city in the dataset, and the result are shown in Fig. \ref{Fig5} of the main text. In addition we have calculate the corresponding quantities for the next 5 most populated cities, ranging from around 500,000 to 1,600,000 inhabitants in 2007. For each city, the shapes of $T_{LCA}$ and $t_{mid}$ are similar to those presented in the main text, showing that the behaviour is a general trait of the people living in urban areas and not a particularity of a specific city, as can be seen in Figs. \ref{SI-Fig1} and \ref{SI-Fig2}.


\begin{figure}[!ht]
\centering
\includegraphics[width=.6\linewidth]{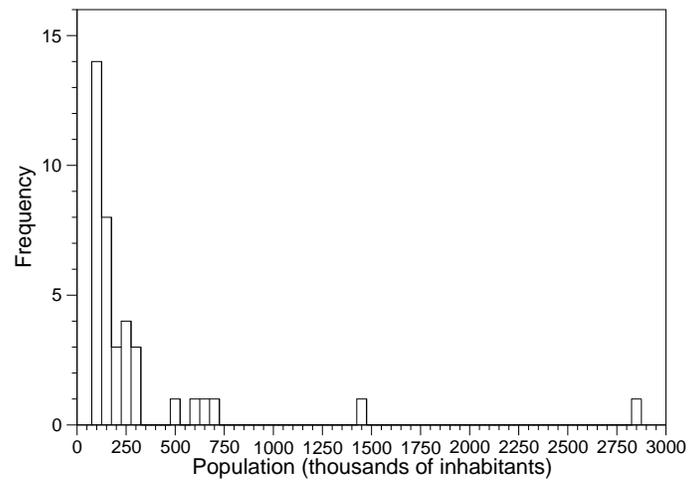}
\caption{{\bf Distribution of cities by population size in 2007.} The values are rounded to multiples of 50,000 to keep the identity of each city unknown.}  
\label{SI-Fig3}
\end{figure}

\begin{thebibliography}{10}

\bibitem{roenneberg2013light}
T.~Roenneberg, T.~Kantermann, M.~Juda, C.~Vetter, and K.~V. Allebrandt, ``Light
  and the human circadian clock,'' in {\em Circadian clocks}, pp.~311--331,
  Springer, 2013.

\bibitem{dinges1995overview}
D.~F. Dinges, ``An overview of sleepiness and accidents,'' {\em Journal of
  sleep research}, vol.~4, no.~s2, pp.~4--14, 1995.

\bibitem{pauley2004lighting}
S.~M. Pauley, ``Lighting for the human circadian clock: recent research
  indicates that lighting has become a public health issue,'' {\em Medical
  hypotheses}, vol.~63, no.~4, pp.~588--596, 2004.

\bibitem{nutt2008sleep}
D.~Nutt, S.~Wilson, and L.~Paterson, ``Sleep disorders as core symptoms of
  depression,'' {\em Dialogues in clinical neuroscience}, vol.~10, no.~3,
  p.~329, 2008.

\bibitem{hidalgo2009relationship}
M.~P. Hidalgo, W.~Caumo, M.~Posser, S.~B. Coccaro, A.~L. Camozzato, and
  M.~L.~F. Chaves, ``Relationship between depressive mood and chronotype in
  healthy subjects,'' {\em Psychiatry and clinical neurosciences}, vol.~63,
  no.~3, pp.~283--290, 2009.

\bibitem{aries2008effect}
M.~B. Aries and G.~R. Newsham, ``Effect of daylight saving time on lighting
  energy use: A literature review,'' {\em Energy policy}, vol.~36, no.~6,
  pp.~1858--1866, 2008.

\bibitem{sampaio2008efficiency}
B.~R. Sampaio, O.~L. Neto, and Y.~Sampaio, ``Efficiency analysis of public
  transport systems: Lessons for institutional planning,'' {\em Transportation
  research part A: policy and practice}, vol.~42, no.~3, pp.~445--454, 2008.

\bibitem{hofstra2008assess}
W.~A. Hofstra and A.~W. de~Weerd, ``How to assess circadian rhythm in humans: a
  review of literature,'' {\em Epilepsy \& Behavior}, vol.~13, no.~3,
  pp.~438--444, 2008.

\bibitem{orzel2010consequences}
J.~Orze{\l}-Gryglewska, ``Consequences of sleep deprivation,'' {\em
  International journal of occupational medicine and environmental health},
  vol.~23, no.~1, pp.~95--114, 2010.

\bibitem{davies2014effect}
S.~K. Davies, J.~E. Ang, V.~L. Revell, B.~Holmes, A.~Mann, F.~P. Robertson,
  N.~Cui, B.~Middleton, K.~Ackermann, M.~Kayser, {\em et~al.}, ``Effect of
  sleep deprivation on the human metabolome,'' {\em Proceedings of the National
  Academy of Sciences}, vol.~111, no.~29, pp.~10761--10766, 2014.

\bibitem{roenneberg2004marker}
T.~Roenneberg, T.~Kuehnle, P.~P. Pramstaller, J.~Ricken, M.~Havel, A.~Guth, and
  M.~Merrow, ``A marker for the end of adolescence,'' {\em Current Biology},
  vol.~14, no.~24, pp.~R1038--R1039, 2004.

\bibitem{roenneberg2007human}
T.~Roenneberg, C.~J. Kumar, and M.~Merrow, ``The human circadian clock entrains
  to sun time,'' {\em Current Biology}, vol.~17, no.~2, pp.~R44--R45, 2007.

\bibitem{roenneberg2007epidemiology}
T.~Roenneberg, T.~Kuehnle, M.~Juda, T.~Kantermann, K.~Allebrandt, M.~Gordijn,
  and M.~Merrow, ``Epidemiology of the human circadian clock,'' {\em Sleep
  medicine reviews}, vol.~11, no.~6, pp.~429--438, 2007.

\bibitem{levandovski2013chronotype}
R.~Levandovski, E.~Sasso, and M.~P. Hidalgo, ``Chronotype: a review of the
  advances, limits and applicability of the main instruments used in the
  literature to assess human phenotype,'' {\em Trends in psychiatry and
  psychotherapy}, vol.~35, no.~1, pp.~3--11, 2013.

\bibitem{horne1975self}
J.~A. Horne and O.~Ostberg, ``A self-assessment questionnaire to determine
  morningness-eveningness in human circadian rhythms.,'' {\em International
  journal of chronobiology}, vol.~4, no.~2, pp.~97--110, 1975.

\bibitem{roenneberg2003life}
T.~Roenneberg, A.~Wirz-Justice, and M.~Merrow, ``Life between clocks: daily
  temporal patterns of human chronotypes,'' {\em Journal of biological
  rhythms}, vol.~18, no.~1, pp.~80--90, 2003.

\bibitem{kovanen2013temporal}
L.~Kovanen, K.~Kaski, J.~Kert{\'e}sz, and J.~Saram{\"a}ki, ``Temporal motifs
  reveal homophily, gender-specific patterns, and group talk in call
  sequences,'' {\em Proceedings of the National Academy of Sciences}, vol.~110,
  no.~45, pp.~18070--18075, 2013.

\bibitem{eagle2009inferring}
N.~Eagle, A.~S. Pentland, and D.~Lazer, ``Inferring friendship network
  structure by using mobile phone data,'' {\em Proceedings of the national
  academy of sciences}, vol.~106, no.~36, pp.~15274--15278, 2009.

\bibitem{jiang2013calling}
Z.-Q. Jiang, W.-J. Xie, M.-X. Li, B.~Podobnik, W.-X. Zhou, and H.~E. Stanley,
  ``Calling patterns in human communication dynamics,'' {\em Proceedings of the
  National Academy of Sciences}, vol.~110, no.~5, pp.~1600--1605, 2013.

\bibitem{blondel2015survey}
V.~D. Blondel, A.~Decuyper, and G.~Krings, ``A survey of results on mobile
  phone datasets analysis,'' {\em EPJ Data Science}, vol.~4, no.~1, pp.~1--55,
  2015.

\bibitem{Bhattacharya160097}
K.~Bhattacharya, A.~Ghosh, D.~Monsivais, R.~I.~M. Dunbar, and K.~Kaski, ``Sex
  differences in social focus across the life cycle in humans,'' {\em Royal
  Society Open Science}, vol.~3, no.~4, 2016.

\bibitem{david2016communication}
T.~David-Barrett, J.~Kertesz, A.~Rotkirch, A.~Ghosh, K.~Bhattacharya,
  D.~Monsivais, and K.~Kaski, ``Communication with family and friends across
  the life course,'' {\em PLOS One}, , vol.~11, no.~11, p.~e0165687, 2016.

\bibitem{torous2015realizing}
J.~Torous, P.~Staples, and J.-P. Onnela, ``Realizing the potential of mobile
  mental health: new methods for new data in psychiatry,'' {\em Current
  psychiatry reports}, vol.~17, no.~8, pp.~1--7, 2015.

\bibitem{song2010limits}
C.~Song, Z.~Qu, N.~Blumm, and A.-L. Barab{\'a}si, ``Limits of predictability in
  human mobility,'' {\em Science}, vol.~327, no.~5968, pp.~1018--1021, 2010.

\bibitem{sevtsuk2010does}
A.~Sevtsuk and C.~Ratti, ``Does urban mobility have a daily routine? learning
  from the aggregate data of mobile networks,'' {\em Journal of Urban
  Technology}, vol.~17, no.~1, pp.~41--60, 2010.

\bibitem{stopczynski2014measuring}
A.~Stopczynski, V.~Sekara, P.~Sapiezynski, A.~Cuttone, M.~M. Madsen, J.~E.
  Larsen, and S.~Lehmann, ``Measuring large-scale social networks with high
  resolution,'' {\em PloS one}, vol.~9, no.~4, p.~e95978, 2014.

\bibitem{jiang2016timegeo}
S.~Jiang, Y.~Yang, S.~Gupta, D.~Veneziano, S.~Athavale, and M.~C. Gonz{\'a}lez,
  ``The timegeo modeling framework for urban motility without travel surveys,''
  {\em Proceedings of the National Academy of Sciences}, p.~201524261, 2016.

\bibitem{sun2013understanding}
L.~Sun, K.~W. Axhausen, D.-H. Lee, and X.~Huang, ``Understanding metropolitan
  patterns of daily encounters,'' {\em Proceedings of the National Academy of
  Sciences}, vol.~110, no.~34, pp.~13774--13779, 2013.

\bibitem{louail2014mobile}
T.~Louail, M.~Lenormand, O.~G.~C. Ros, M.~Picornell, R.~Herranz,
  E.~Frias-Martinez, J.~J. Ramasco, and M.~Barthelemy, ``From mobile phone data
  to the spatial structure of cities,'' {\em Scientific reports}, vol.~4, 2014.

\bibitem{abdullah2014towards}
S.~Abdullah, M.~Matthews, E.~L. Murnane, G.~Gay, and T.~Choudhury, ``Towards
  circadian computing: early to bed and early to rise makes some of us
  unhealthy and sleep deprived,'' in {\em Proceedings of the 2014 ACM
  international joint conference on pervasive and ubiquitous computing},
  pp.~673--684, ACM, 2014.

\bibitem{aledavood2015daily}
T.~Aledavood, E.~L{\'o}pez, S.~G. Roberts, F.~Reed-Tsochas, E.~Moro, R.~I.
  Dunbar, and J.~Saram{\"a}ki, ``Daily rhythms in mobile telephone
  communication,'' {\em PloS one}, vol.~10, no.~9, p.~e0138098, 2015.

\bibitem{aledavood2016channel}
T.~Aledavood, E.~L{\'o}pez, S.~G. Roberts, F.~Reed-Tsochas, E.~Moro, R.~I.
  Dunbar, and J.~Saram{\"a}ki, ``Channel-specific daily patterns in mobile
  phone communication,'' in {\em Proceedings of ECCS 2014}, pp.~209--218,
  Springer, 2016.

\bibitem{monsivais2017seasonal}
D.~Monsivais, K.~Bhattacharya, A.~Ghosh, and K.~Kaski, ``Seasonal and geographical impact on human resting periods,'' {\em Scientific Reports}, vol.~7, 2017.

\bibitem{christensen2016direct}
M.~A. Christensen, L.~Bettencourt, L.~Kaye, S.~T. Moturu, K.~T. Nguyen, J.~E.
  Olgin, M.~J. Pletcher, and G.~M. Marcus, ``Direct measurements of smartphone
  screen-time: Relationships with demographics and sleep,'' {\em PloS one},
  vol.~11, no.~11, p.~e0165331, 2016.

\bibitem{cuttone2017sensiblesleep}
A.~Cuttone, P.~B{\ae}kgaard, V.~Sekara, H.~Jonsson, J.~E. Larsen, and
  S.~Lehmann, ``Sensiblesleep: A bayesian model for learning sleep patterns
  from smartphone events,'' {\em PloS one}, vol.~12, no.~1, p.~e0169901, 2017.

\bibitem{mollgaard2016general}
A.~Mollgaard, S.~Lehmann, and J.~Mathiesen, ``General human activity
  patterns,'' {\em arXiv preprint arXiv:1611.08262}, 2016.

\bibitem{duffy1999relationship}
J.~Duffy, D.-J. Dijk, E.~F. Hall, and C.~A. Czeisler, ``Relationship of
  endogenous circadian melatonin and temperature rhythms to self-reported
  preference for morning or evening activity in young and older people,'' {\em
  health}, vol.~26, p.~34, 1999.

\bibitem{dijk1997variation}
D.-J. Dijk, T.~L. Shanahan, J.~F. Duffy, J.~M. Ronda, and C.~A. Czeisler,
  ``Variation of electroencephalographic activity during non-rapid eye movement
  and rapid eye movement sleep with phase of circadian melatonin rhythm in
  humans,'' {\em The Journal of Physiology}, vol.~505, no.~3, pp.~851--858,
  1997.

\bibitem{dijk2012amplitude}
D.-J. Dijk, J.~F. Duffy, E.~J. Silva, T.~L. Shanahan, D.~B. Boivin, and C.~A.
  Czeisler, ``Amplitude reduction and phase shifts of melatonin, cortisol and
  other circadian rhythms after a gradual advance of sleep and light exposure
  in humans,'' {\em PloS one}, vol.~7, no.~2, p.~e30037, 2012.

\bibitem{allebrandt2014chronotype}
K.~V. Allebrandt, M.~Teder-Laving, T.~Kantermann, A.~Peters, H.~Campbell,
  I.~Rudan, J.~F. Wilson, A.~Metspalu, and T.~Roenneberg, ``Chronotype and
  sleep duration: the influence of season of assessment,'' {\em Chronobiology
  international}, vol.~31, no.~5, pp.~731--740, 2014.

\bibitem{foster2008human}
R.~G. Foster and T.~Roenneberg, ``Human responses to the geophysical daily,
  annual and lunar cycles,'' {\em Current biology}, vol.~18, no.~17,
  pp.~R784--R794, 2008.
  
\bibitem{ungeo}
Standard country or area codes for statistical use. {\url{https://unstats.un.org/unsd/methodology/m49/}}. Accessed: 2017-09-28.  

\bibitem{lockley2012sleep}
S.~W. Lockley and R.~G. Foster, {\em Sleep: a very short introduction},
  vol.~295.
\newblock Oxford University Press, 2012.

\bibitem{wittmann2006social}
M.~Wittmann, J.~Dinich, M.~Merrow, and T.~Roenneberg, ``Social jetlag:
  misalignment of biological and social time,'' {\em Chronobiology
  international}, vol.~23, no.~1-2, pp.~497--509, 2006.

\bibitem{kullback1951information}
S.~Kullback and R.~A. Leibler, ``On information and sufficiency,'' {\em The
  annals of mathematical statistics}, vol.~22, no.~1, pp.~79--86, 1951.

\end{thebibliography}
\end{document}